\documentclass[aps,prl,reprint,groupedaddress,amsmath,amssymb,showpacs,floatfix]{revtex4-1}
\usepackage{graphicx}
\usepackage{natbib}
\usepackage{upgreek}
\usepackage{textcomp}
\usepackage[colorlinks,linkcolor=blue,citecolor=blue]{hyperref}
\usepackage{color}

\begin{document}

%Title of paper
\title{Highly-sensitive superconducting quantum interference proximity transistor}

\author{Alberto Ronzani}
%\email{alberto.ronzani@nano.cnr.it}
\affiliation{NEST, Istituto Nanoscienze-CNR and Scuola Normale Superiore, 
             I-56127 Pisa, Italy}
	     
\author{Carles Altimiras}
\email{carles.altimiras@sns.it}
\affiliation{NEST, Istituto Nanoscienze-CNR and Scuola Normale Superiore, 
             I-56127 Pisa, Italy}

\author{Francesco Giazotto}
%\email{francesco.giazotto@sns.it}
\affiliation{NEST, Istituto Nanoscienze-CNR and Scuola Normale Superiore, 
             I-56127 Pisa, Italy}

\begin{abstract}
	We report the design and implementation of a high-performance
	superconducting quantum interference proximity transistor (SQUIPT)
	based on aluminum-copper (Al-Cu) technology. 
        With the adoption of a thin and short copper nanowire
	we demostrate full phase-driven modulation of the proximity-induced minigap in the normal metal
	density of states.
	Under optimal bias we record unprecedently high flux-to-voltage (up to $3\,\mathrm{mV}/\Phi_0$)
	and flux-to-current (exceeding $100\,\mathrm{nA}/\Phi_0$) transfer function
	values at sub-Kelvin temperatures, where $\Phi_0$ is the flux quantum.
	The best magnetic flux resolution (as low as 
	$500\,\mathrm{n}\Phi_0/\sqrt{\mathrm{Hz}}$ at 240 mK, being limited by the
	room temperature pre-amplification stage) is reached under fixed current bias.
	These figures of merit combined with ultra-low power dissipation and micrometer-size dimensions
	make this mesoscopic interferometer attractive for low-temperature
	applications such as the investigation of the magnetization of small spin populations.
\end{abstract}


\maketitle

\begin{figure}
\begin{center}
	\includegraphics[width=3.26in]{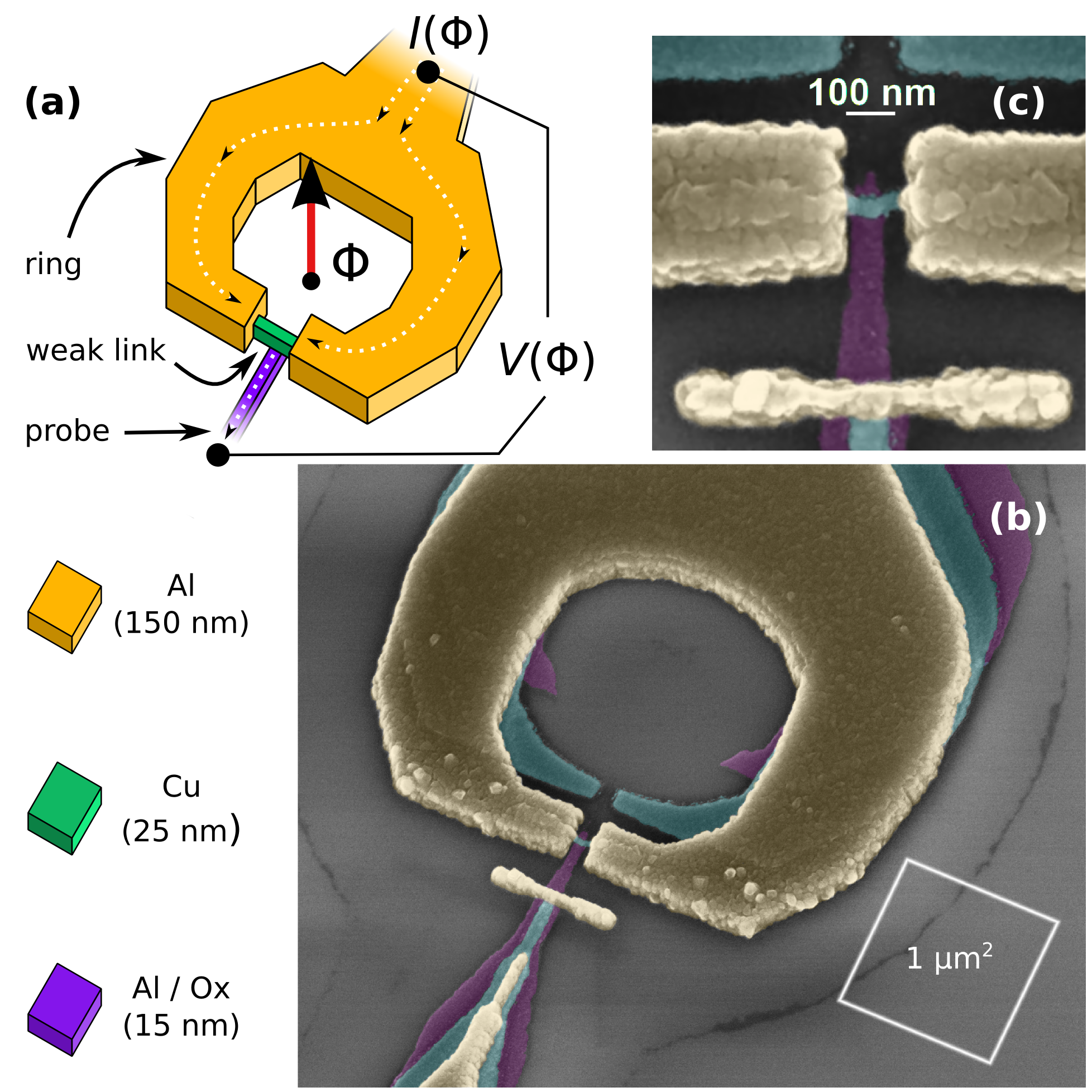}
\end{center}
\caption{
\textbf{(a)} Functional schematic of a SQUIPT device,
realized by tilted evaporation of metallic thin films through a suspended
resist mask defined by electron-beam lithography. 
The first evaporation (purple layer) consists in $15\,\mathrm{nm}$
of aluminum (Al), subsequently oxidized to form an AlOx tunnel barrier. 
The second evaporation (green layer) consists in $25\,\mathrm{nm}$ of
copper (Cu) realizing the normal metal nanowire. 
Finally, $150\,\mathrm{nm}$ of Al (dark yellow) are evaporated to form the superconducting
loop (having inner diameter $\simeq 1.7\, \upmu\mathrm{m}$) as well as the electrodes
in clean electric contact with the Cu film. In the schematic $\Phi$ is the magnetic flux linked to the
superconducting loop, $I(\Phi)$ and $V(\Phi)$ are respectively the current flowing through and the
voltage difference across the device.
\textbf{(b)} Tilted scanning electron micrograph showing the complete SQUIPT loop.
\textbf{(c)} Scanning electron micrograph centered on the Cu nanowire region.
The interelectrode spacing is $\simeq 140\, \mathrm{nm}$;
 the width of the nanowire is $ \simeq 30\, \mathrm{nm}$.
The tunnel probe is $\simeq 60\,\mathrm{nm}$-wide at the interface with the Cu weak-link.
}
\label{fig:fig1}
\end{figure}

The superconducting quantum interference proximity transistor \cite{giazotto2010superconducting}
(SQUIPT) is a two terminal device based on a tunnel junction between a superconducting
``probe'' electrode and a normal metal wire inserted in a superconducting
loop, see Fig. \ref{fig:fig1}(a) .
Owing to \textit{proximity effect} \cite{degennes1966superconductivity}
the clean electric contact between the normal and the superconducting
metal leads to the appearance of a phase-dependent
minigap in the electronic density of states (DOS) of the normal metal \cite{lesueur2008phase}.
Yet, once proximized, the latter estabilishes a
weak coupling \cite{josephson1962possible} between the superconducting electrodes in contact with it, so that the
whole superconductor-normal metal-superconductor (SNS) complex operates as a Josephson junction \cite{likharev1979superconducting}.
An external magnetic field threading the loop therefore modulates
the amplitude of the minigap by establishing a definite phase difference
across the Josephson junction as a consequence of flux quantization in the
closed superconducting ring \cite{doll1961experimental,deaver1961experimental}.

SNS weak-links offer the possibility to realize superconducting couplings unachievable by Josephson
tunnel junctions, for instance, providing non sinusoidal current-phase relation in the short junction
 length limit \cite{likharev1979superconducting}, or achieving a $\pi$-state by driving the weak-link 
into non-equilibrium conditions \cite{baselmans2002direct}. In SQUIPTs, the SNS weak-link is exploited so that the electronic DOS of the metallic wire is modulated \cite{petrashov1994conductivity,petrashov1995phase,lesueur2008phase} 
by externally driving the magnetic flux. 
Therefore SQUIPTs can be considered as a magnetic analog 
of semiconductor field effect transistors, and can be exploited as sensitive magnetic flux sensors. 
Being based on mesoscopic quantum effects, they bear technological similarities 
with Single Electron Transistors (SETs) used as charge sensors \cite{devoret2000amplifying}: 
they require being operated at sub-Kelvin temperatures and are characterized by moderately high output impedances 
limiting their bandwidth unless radiofrequency matching techniques are used \cite{schoelkopf1998radiofrequency}. 
Despite these limitations, they are well worth the effort since their performance 
in terms of flux sensitivity and power dissipation has no technological counterpart. 
For instance, ultimate magnetic flux resolution values as low as $1\, \mathrm{n\Phi_0/\sqrt{Hz}}$ below 1K
(where $\Phi_0 = h/2e \simeq 2.067 \times 10^{-15} \, \mathrm{Wb}$ is the flux quantum)
have been predicted for SQUIPTs based on large energy-gap superconducting materials such as vanadium \cite{giazotto2011hybrid}.
This makes them attractive for low-temperature applications such as the investigation of the magnetization
of small spin populations.

Here we report the realization and characterization of a SQUIPT based on the Al-Cu technology
which shows record flux-to-voltage transfer function values (up to $\simeq 3\, \mathrm{mV/\Phi_0}$), 
leading to a magnetic flux resolution $\Phi_{N} \simeq 500 \, \mathrm{n\Phi_0/\sqrt{Hz}}$ at $240\,\mathrm{mK}$
in the near DC frequency range and being limited by the voltage noise of the room temperature 
preamplification stage. 

The earliest proof-of-concept realization \cite{giazotto2010superconducting} of a SQUIPT described a device based on 
a copper wire of length~$L=1.5\, \upmu\mathrm{m}$ proximized by a thin aluminum loop
spanning a surface of~$\approx 120\, \upmu\mathrm{m}^2$. These figures resulted
in the SNS junction being in the \textit{long} diffusive limit (\textit{i.e.}, $L >> \xi_0$, where $\xi_0$ is the coherence length),
 entailing modest minigap modulation
amplitudes~($\approx 10\,\upmu\mathrm{eV}$) under optimal current bias.
In subsequent attempts aiming at increasing the response of the device up to its predicted
intrinsic limits \cite{giazotto2011hybrid}, the length of the N wire
has been reduced to bring the SNS junction in the \textit{short} regime, where the
amplitude of the proximized minigap can approach that of the ``parent'' superconductor.
These devices \cite{meschke2011tunnel} succeeded in achieving wider induced minigap values~($\approx 130\,\upmu\mathrm{eV}$),
but suffered from hysteresis stemming from self-induced magnetic screening
caused by the high critical current magnitude typical of short metallic weak-links.
This shortcoming has recently been lifted by reducing the cross-section of the copper wire
while keeping it in the short-junction limit ($L\approx 250\, \mathrm{nm}$)
therefore allowing to reach a sizeable magnetic flux responsivity, with an estimated~$\approx 50\, \upmu\mathrm{eV}$ 
minigap modulation amplitude, \textit{i.e.}, $\approx 36 \%$ of the full induced minigap width
deduced from the differential conductance measurements \cite{jabdaraghi2014nonhysteretic}.

The cause for incomplete minigap modulation lies in non-ideal phase bias
of the weak-link. In the limit of neglegible magnetic screening originating from the geometric self-inductance
of the loop, 
the effective phase difference imposed to the weak-link by flux quantization is affected
by the competition between the kinetic inductance of the superconducting loop and that 
of the weak-link (respectively $\mathcal{L}^S$ and $\mathcal{L}^{WL}$). Complete phase-bias,
corresponding to high phase gradient developed across the Cu nanowire, 
is only possible in the limit $\mathcal{L}^S/\mathcal{L}^{WL} \ll 1$ \cite{lesueur2008phase}.
The difficulty in achieving this regime originates from the short-junction nature of the weak-links which, apart from the 
aforementioned high critical current values, also show at low temperature
a non-sinusoidal current-phase relationship $I^{WL}(\phi)$.
Both effects \cite{likharev1979superconducting,golubov2004currentphase}
suppress the magnitude of $\mathcal{L}^{WL}= (\Phi_0/2\pi)(\partial I^{WL} / \partial \phi)^{-1}$
as the value of the phase difference $\phi$ approaches $\pi$,
where the sharpest response is expected \cite{giazotto2011hybrid}.
In the present SQUIPT we show that a complete phase
bias can be achieved in a junction approaching the short limit by realizing a copper wire having a nanoscale cross-section
(thus maximizing $\mathcal{L}^{WL}$) while at the same time having a compact superconducting aluminum loop characterized
by low normal-state resistance (therefore minimizing both the kinetic and geometric components of $\mathcal{L}^{S}$).
As a consequence of the full minigap modulation in the proximized weak-link,
we obtain record magnetic flux responsivity figures, both in current and voltage biased setups.

Figure~\ref{fig:fig1}(b) shows the scanning electron micrograph of a typical SQUIPT device, 
fabricated by standard electron-beam lithography on a suspended bilayer resist mask \cite{dolan1977offset}
($1000\,\mathrm{nm}$ copolymer / $100\, \mathrm{nm}$ polymethyl-methacrylate) on top of
an oxidized silicon substrate. An initial $15$-nm-thick Al layer is deposited
at $40^{\circ}$ via electron-beam evaporation in ultra-high vacuum conditions ($\simeq 10^{-9}\,\mathrm{Torr}$),
and subsequently exposed to a pure $\mathrm{O}_2$ atmosphere ($37\,\mathrm{mTorr}$ for $300\,\mathrm{s}$) to obtain
the tunnel probe electrode. The normal metal nanowire is realized by evaporating 
a $25$-nm-thick Cu layer at $20^{\circ}$. Finally, a $150$-nm-thick Al film in clean contact
with the latter layer is deposited at zero angle to implement the superconducting loop, 
designed to have an internal diameter~$\simeq 1.7\,\upmu\mathrm{m}$. The device core [see 
Fig.~\ref{fig:fig1}(c)] shows an interelectrode spacing $L \simeq 140\,\mathrm{nm}$,
while the copper nanowire is $30\,\mathrm{nm}$-wide and overlaps the lateral superconducting electrodes
for $\simeq 400\, \mathrm{nm}$ per side.
The width of the tunnel probe electrode is $\simeq 60\, \mathrm{nm}$.
Based on previous measurement on Cu nanowires of similar cross-section \cite{ronzani2014balanced},
we estimate the ratio $L/\xi_0 = L \sqrt{ \Delta_r / \hbar D_{Cu} } \approx 1.1$ which confirms
the frame of the intermediate-short junction regime. Above, $\Delta_r \simeq 185\, \upmu\mathrm{eV}$ is the 
energy gap in the superconducting ring and $D_{Cu} \simeq 55\, \mathrm{cm^2/s}$ is the diffusion coefficient for
our Cu nanowire. The yield of such simple fabrication scheme is about 20\%, being mostly limited by the mechanical stability of the suspended polymethyl-methacrylate mask defining the loop.

The SQUIPT magneto-electric characterization was performed in a filtered ${}^3\mathrm{He}$
cryostat for several temperatures in the range $0.24 - 0.85\,\mathrm{K}$. Current response under voltage
bias was measured in a two-wire configuration with a commercial current preamplifier
(DL Instruments model 1211) as a function of the
magnetic flux generated by a magnetic field applied orthogonally to the plane of the substrate. The current response shows periodicity
with respect to the applied magnetic field density with period $B_0 = \Phi_0/A_{eff} \approx 6.2\,\mathrm{G}$,
where $A_{eff} \approx 3.3\,\upmu\mathrm{m}^2$ is consistent with the area enclosed by the ring of the SQUIPT.

\begin{figure}
\begin{center}
	\includegraphics[width=3.26in]{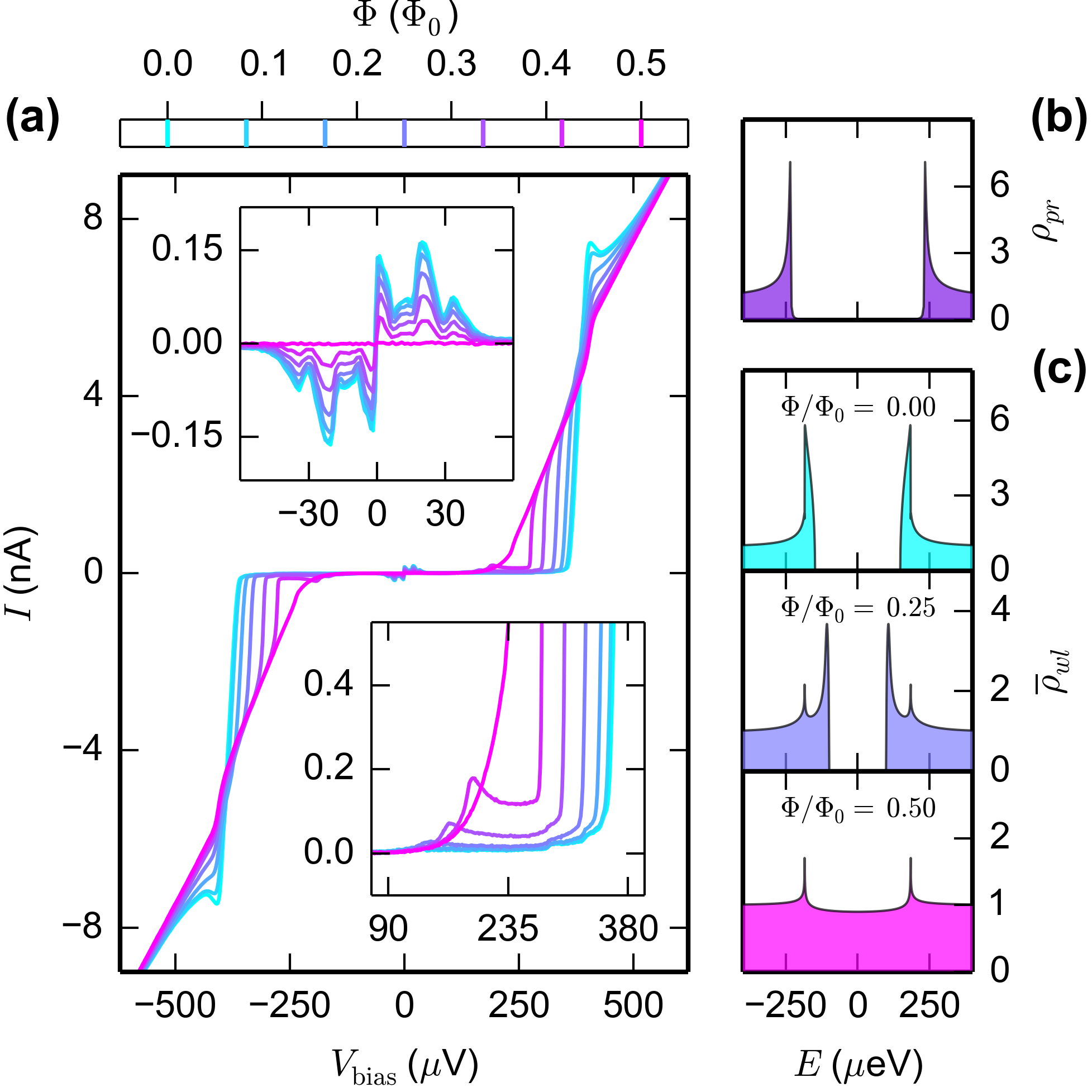}
\end{center}
\caption{
\textbf{(a)} SQUIPT current vs voltage characteristics recorded at $240 \, \mathrm{mK}$ 
for 7 equally-spaced magnetic flux values ranging from $\Phi=0$ to $\Phi=\Phi_0/2$.
The upper inset shows a magnification of the low-voltage bias range, where weak
phase-dependent supercurrent features appear at fixed voltage values.
The lower inset shows a magnification of the onset of quasiparticle conduction, where the voltage
dependence of the current can be non-monotonic as a consequence of thermally-activated transport.
This is particularly evident (at finite temperature) when the minigap starts to be suppressed by the magnetic flux.
\textbf{(b)} Theoretical BCS density of states $\rho_{pr}(E)$ of the superconducting probe 
($\Delta_{pr} = 235\,\upmu\mathrm{eV}$).
\textbf{(c)}  Theoretical local density of states $\overline{\rho}_{wl}(E)$
 in the proximized weak-link averaged
over the probe width for three different values of the applied magnetic flux. $\overline{\rho}_{wl}(E)$ was
obtained by the numerical solution of the 1-D Usadel equations assuming $L = 1.1\,\xi_0$, and full
transparency at the interfaces.
}
\label{fig:fig2}
\end{figure}

Figure~\ref{fig:fig2}(a) shows the current vs voltage $I(\mathrm{V_{bias}})$
characteristics recorded at the base temperature ($T = 240\,\mathrm{mK}$),
for selected equally-spaced values of the applied magnetic flux ranging from $\Phi=0$ to $\Phi=\Phi_0/2$.
At zero flux (fully open minigap) the characteristic shows a behaviour resembling that of a tunnel junction
between superconductors with different energy gaps. By increasing the magnetic field
the minigap closes, until the characteristic is similar to that of a $NIS$ junction at $\Phi=\Phi_0/2$.
From these data we estimate the $15$-nm-thick aluminum probe to be characterized by a superconducting gap 
$\Delta_{pr}\approx 235\,\upmu\mathrm{eV}$ 
and a tunnel resistance $R_T \approx 55\, \mathrm{k\Omega}$. The curves are consistent with a maximum
minigap amplitude $\varepsilon_g(\Phi=0) \approx 145\, \upmu\mathrm{eV}$, 
a value which corresponds approximately
to $78\%$ of the energy gap in the superconducting ring. 
Figure~\ref{fig:fig2}(b) shows the theoretical 
Bardeen-Cooper-Schrieffer (BCS)-like profile \cite{bardeen1957theory} of the DOS in the probe junction 
$\rho_{pr}(E)=\left|\mathrm{Re}(\frac{E+i\gamma}{\sqrt{(E+i\gamma)^2-\Delta_{pr}^2}})\right|$ 
where $\gamma/\Delta_{pr}=10^{-3}$ accounts for energy smearing due to finite quasiparticle lifetime \cite{dynes1978direct, pekola2010environmentassisted}. 
Panels in figure~\ref{fig:fig2}(c)
show the theoretical DOS in the weak link $\overline{\rho}_{WL}(E)$, spatially averaged over the probe width.
The latter DOS has been obtained by solving numerically the 1-D Usadel equations \cite{virtanen2007thermoelectric, hammer2007density} 
with parameters  $L/\xi_0 = 1.1$, $\Delta_r = 185\, \upmu\mathrm{eV}$ and assuming perfect interface 
transmissivity between the ring and the wire, for $\Phi=\{0, 0.25, 0.5\}\Phi_0$. 

The upper inset of Fig.~\ref{fig:fig2}(a) shows a magnified view of the flux dependent features appearing at low bias. The latter 
can be attributed to a weak Josephson coupling between the proximized nanowire and the probe electrode, 
and their complete suppression at $\Phi = \Phi_0/2$ is a further indication of the full modulation of the
minigap.  
A close inspection of the current vs voltage characteristics in Fig.~\ref{fig:fig2}(a) indicates that
the measured current modulation is able to reach peak-to-peak amplitudes
 as large as $4\, \mathrm{nA}$. On the other hand, when biased at fixed current, the amplitude 
of the corresponding voltage modulation approaches $\varepsilon_g/e$. 

The lower inset in Fig.~\ref{fig:fig2}(a) shows a blow-up of the characteristics at the onset
of quasiparticle conduction. Here the current is non monotonic as a consequence of 
the appreciable thermal population of the quasiparticle states in the proximized nanowire
resulting in additional conduction when 
$\mathrm{V_{bias}} = [\Delta_{pr} - \varepsilon_{g}(\Phi)]/e$.
This bias configuration shifts the chemical potentials of the tunnel junction electrodes 
so that the singularity in the empty DOS of the probe electrode is energetically aligned to
the thermally excited quasiparticles in the copper nanowire.
In the following, we adopt the term ``singularity-matching peak'' to refer to this particular
transport feature, in analogy to $\mathrm{S_1 I S_2}$ systems~\cite{tinkham2004introduction}.
The tunnel resistance value obtained in the fabrication process
is compatible with the optimal input load impedance of both voltage and current preamplifiers.
In the following we consider both voltage-biased and current-biased setups.

\begin{figure}
\begin{center}
	\includegraphics[width=3.26in]{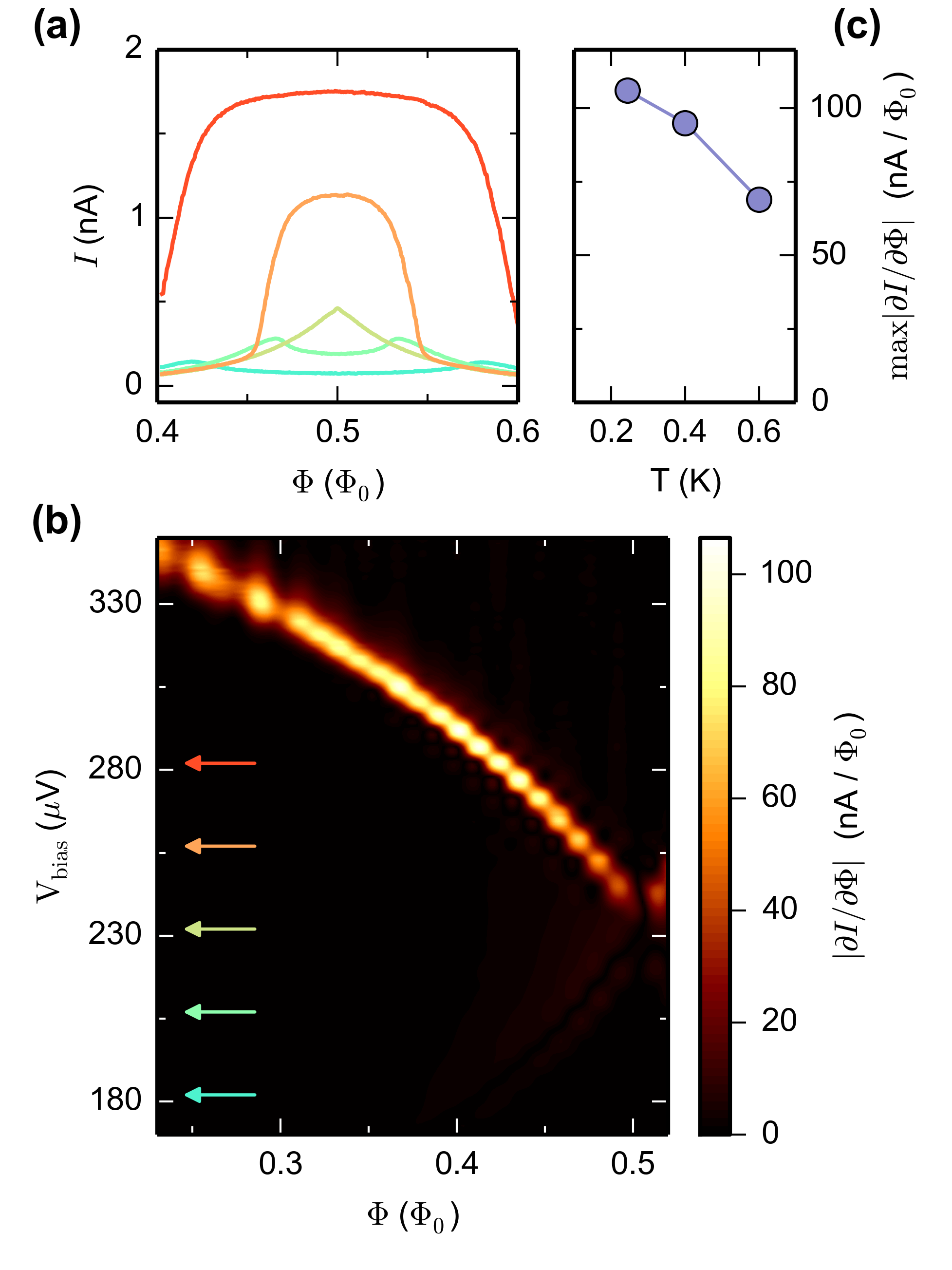}
\end{center}
\caption{
\textbf{(a)} Current vs flux characteristics measured at $240\, \mathrm{mK}$ for fixed voltage bias
(from bottom to top, $\mathrm{V_{bias}} = 182,\, 212,\, 232,\, 252,\, 282\, \upmu\mathrm{V}$).
\textbf{(b)} Color plot of the absolute value of the flux-to-current transfer function 
($|\partial I / \partial \Phi|$ vs $\mathrm{V_{bias}}$ and $\Phi$)
obtained by numerical differentiation of the $I(\Phi)$ curves measured at $240\, \mathrm{mK}$.
Arrows indicate in corresponding colors the voltage bias values for the characteristics plotted in panel (a).
\textbf{(c)} Temperature dependence of the maximum absolute value of the flux-to-current transfer function. 
}
\label{fig:fig3}
\end{figure}

\begin{figure*}
\begin{center}
	\includegraphics[width=6.67in]{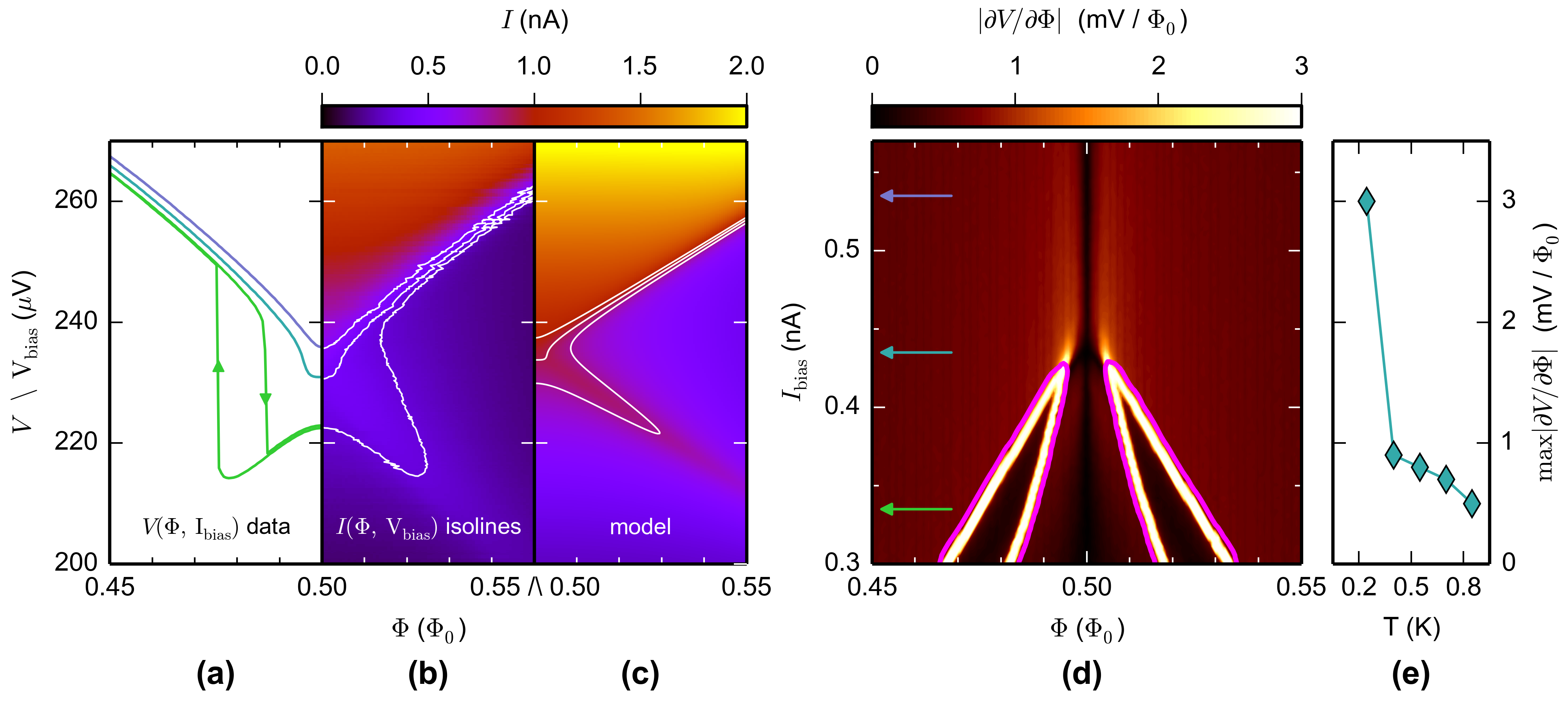}
\end{center}
\caption{
\textbf{(a)} Voltage vs flux characteristics measured at $240\, \mathrm{mK}$ for fixed current bias
(from bottom to top, $\mathrm{I_{bias}} = 335,\, 435,\, 535\, \mathrm{pA}$).
\textbf{(b)} Color plot showing the measured  $I(\mathrm{V_{bias}},\,\Phi$) dataset;
 the current isolines match the characteristics shown in panel (a). The reentrant shape of the
lowest current isoline is at the origin of the hysteresis displayed in the corresponding $V(\Phi,\,\mathrm{I_{bias}})$ characteristic [green trace in panel (a)].
\textbf{(c)} Color plot of the theoretical current vs $\mathrm{V_{bias}}$ and $\Phi$ calculated for a
SQUIPT device based on a weak-link in the short regime.
\textbf{(d)} Color plot of the absolute
flux-to-voltage transfer function ($|\partial V / \partial \Phi|$ vs $\mathrm{V_{bias}}$ and $\Phi$)
obtained by numerical differentiation of $V(\Phi)$ curves measured at $240\, \mathrm{mK}$.
Hysteresis originating from the non-monotonicity of the current vs voltage characteristics 
[see lower inset of Fig.~\ref{fig:fig2}(a)] can be appreciated for $I_{\mathrm{bias}} < 435\, \mathrm{pA}$.
The associated switching events are marked by green arrows in panel (a). Magenta lines mark the hysteretic
regions in the colormap. Arrows indicate in corresponding colors the current bias values for the
characteristics plotted in panel (a).
\textbf{(e)} Temperature dependence of the maximum absolute value of the flux-to-voltage transfer function. 
}
\label{fig:fig4}
\end{figure*}

Current vs flux (\textit{i.e.}, at fixed voltage bias) response figures
 have been obtained by 
numerical differentiation with respect to the magnetic flux of the $I(\mathrm{\Phi, \, V_{bias}})$
 characteristics [Figure~\ref{fig:fig3}(a)] at fixed $\mathrm{V_{bias}}$.
Figure~\ref{fig:fig3}(b) shows the absolute value of the base temperature flux-to-current transfer
function $| \partial I (\mathrm{V_{bias}}, \Phi) / \partial \Phi| $ as a colormap.
In this context the transfer function map indicates sharp response (due to the abrupt onset of quasiparticle
conduction) reaching values as high as $108\,\mathrm{nA}/\Phi_0$, 
over a wide range of the bias parameters in both flux: $\Phi \in [0.35 \div 0.45\,\Phi_0]$, 
and voltage bias: $\mathrm{V_{bias}} \in [275 \div 310\,\upmu\mathrm{V}]$.
 This high sensitivity (approximately four times higher than
the best-performing devices so far \cite{jabdaraghi2014nonhysteretic})
originates from both a lower tunnel probe resistance
and a full modulation of the minigap. In addition, the maximum flux-to-current responsivity level, 
is only moderately suppressed by increasing the temperature [see Fig.~\ref{fig:fig3}].

Voltage vs flux characteristics, recorded at $240\,\mathrm{mK}$
 for a few selected values of the current bias 
in the vicinity of the maximal response ($\mathrm{I_{bias}} =435\,\mathrm{pA}$, 
$\Phi\simeq 0.5\,\Phi_0$) are shown  
in Fig.~\ref{fig:fig4}(a). The absolute value of the relative flux-to-voltage transfer function is plotted
as a colormap in Fig.~\ref{fig:fig4}(d). At low current bias the non-monotonicity of the current vs voltage characteristics [see the bottom inset of Fig.~\ref{fig:fig2}(a)] results into
bistable voltage configurations, giving rise to hysteretic behaviour 
and limiting the useful bias range for a SQUIPT used as a linear sensor. 
The non-monotonicity originating from the singularity-matching peak can, in principle,
be limited by lowering the electron temperature beyond the base temperature of 
our cryostat.
This can be achieved by using dilution refrigerators but also with the adoption of 
integrated on-chip electronic coolers \cite{giazotto2006opportunities,nguyen2013trapping}
relying on the same fabrication technique. 
On the other side, the supercurrent peaks give rise to a similar
electric bistability [see Fig.~\ref{fig:fig2}(a), top inset]
and are expected to increase in magnitude at lower temperatures
(when not countered by lower transparency of the tunnel barrier) 
and will ultimately limit the current bias range
available for linear response. 

 Notably, the electric bistability provided by the singularity-matching peak
 could instead be exploited for operating the SQUIPT as a threshold detector.
In this configuration, the flux is applied in the close vicinity of a switching point 
[\textit{e.g.}, $\mathrm{I_{bias}}=335\,\mathrm{pA}$ and $\Phi=0.48\,\Phi_0$ in Fig.~\ref{fig:fig4}(d)], 
so that flux variations crossing the threshold given by the swiching point yield 
a voltage step response (whose amplitude may be $\approx 50\,\upmu\mathrm{V}$) 
within a timescale corresponding to the relaxation time of the measurement setup. 
Such scheme can be useful for sampling the probability distribution function of a noisy magnetic
 flux source. 
%for instance the skewness of its statistics can be extracted by comparing the outcomes 
%corresponding to flux polarizations symmetric with respect to $0.5\Phi_0$.

We now discuss the SQUIPT sensitivity when operated as a linear flux sensor. Inspection of the flux-to-voltage transfer function [shown as a colormap in Fig.~\ref{fig:fig4}(d)] reveals
that the current-biased setup allows for a high responsivity ($\approx 1\, \mathrm{mV}/\Phi_0$)
over a rather broad flux and current bias range, with a peak value of $\approx 3\, \mathrm{mV}/\Phi_0$,
located at $\Phi=0.495\, \Phi_0$ and $\mathrm{I_{bias}} = 435\,\mathrm{pA}$. This working point
lies just outside of the hysteretic region [marked with magenta lines in Fig.~\ref{fig:fig4}(d)] and
it is thus a suitable point for linear operation of the detector.

While compatible with earlier results \cite{jabdaraghi2014nonhysteretic}, 
such a high voltage transfer function may seem surprising.
Indeed, the detailed theoretical investigation of the SQUIPT performance in the short junction limit
(where analytic calculations can be performed) carried in \cite{giazotto2011hybrid}, 
predicted maximal transfer functions of about $3.1\Delta_{r}/e\Phi_0$ around
$\Phi=0.5\,\Phi_0$, which can be traced back to the flux dependence of the minigap.
Extrapolating this limit to our moderately-short SNS junction, \textit{i.e.},
assuming the same scaling but replacing $\Delta_r$ with
the measured minigap $\varepsilon_{g}=145\,\upmu\mathrm{eV}$, one would expect a response of 
about $450\,\upmu\mathrm{eV}/\Phi_0$, which is 6 times smaller than the maximum value obtained in the 
experiment.
The reason for the observed higher response stems from the contribution of the singularity-matching peak, 
ignored in \cite{giazotto2011hybrid}, which bends the non-hysteretic $V(\Phi, \mathrm{I_{bias}})$ 
characteristics in the vicinity of $\Phi_0/2$ and $V=\Delta_{pr}/e$, 
resulting in a sharper voltage response. 
This feature can be easily reproduced by using a simplified model which holds in the
short-junction limit as described in \cite{giazotto2011hybrid},
with the replacement $\Delta_{r} \leftrightarrow \varepsilon_{g}$.

Figure~\ref{fig:fig4}(c) presents a contour plot of the $I(\Phi,\,\mathrm{V_{bias}})$ dataset
 obtained with the above theoretical model in the vicinity of 
$e\mathrm{V_{bias}}=\Delta_{pr}=235\,\upmu\mathrm{eV}$ and $\Phi/\Phi_0=0.5$ at $T=240\,\mathrm{mK}$. 
The white lines correspond to calculated current isolines, who strongly resemble those 
observed in the actual measured $I(\Phi, \, \mathrm{V_{bias}})$ characteristics [see Fig.~\ref{fig:fig4}(b)].
Although rather idealized, the model provides a satisfactory reproduction of the physical mechanism
underlying the observed high responsivity.
Furthermore, a close match between the measured $V(\Phi,\,\mathrm{I_{bias}})$ response curves
 and the current isolines
can be appreciated in the juxtaposition of panels (a,b) in Fig.~\ref{fig:fig4}, 
therefore corroborating the identification of physical origin of the high flux-to-voltage responsivity we
observe. 

In particular, three regimes can be recognized, depending on the magnitude of the quasiparticle current.
Low-current regime [corresponding to the green trace and arrow in Fig.~\ref{fig:fig4}(a,d)] is characterized
by hysteresis originating from the singularity-matching peak bistability, which is evident in the reentrant
shape of the low-current isolines in Fig.~\ref{fig:fig4}(b,c). Conversely, in the high-current limit
[exemplified by the blue trace and arrow in Fig.~\ref{fig:fig4}(a,d)] no hysteresis can be found, but
the magnetic flux responsivity is only moderate. The optimal regime for sensitivity
[represented by the cyan trace and arrow in Fig.~\ref{fig:fig4}(a,d)] emerges in the smooth transition
between the two abovementioned limits. This latter regime features the highest value of the flux-to-voltage
transfer function, but no hysteresis.

The temperature dependence of the maximum value of the transfer function is displayed in Fig.~\ref{fig:fig4}(e).
The substantial enhancement observed at lower temperature is due to the abrupt character of the thermal suppression 
of the singularity-matching peak appearing in the current vs voltage characteristics, which
allows to access the optimal current-bias range required for the sharpest voltage response.

Our device has been designed to show that high transfer function values can be obtained 
in SQUIPTs based on Al-Cu technology. The intermediate value of the impedance of the device 
($R_T = 55\, \mathrm{k\Omega}$), allows sensitive operation of the interferometer in both voltage-biased
and current-biased setups. However, given the significant capacitive load present in the filtered
lines of our refrigerator setup, the current-biased measurement scheme shows better performance
thanks to the superior common-mode noise rejection properties of differential voltage preamplification.

\begin{figure}
\begin{center}
	\includegraphics[width=3.26in]{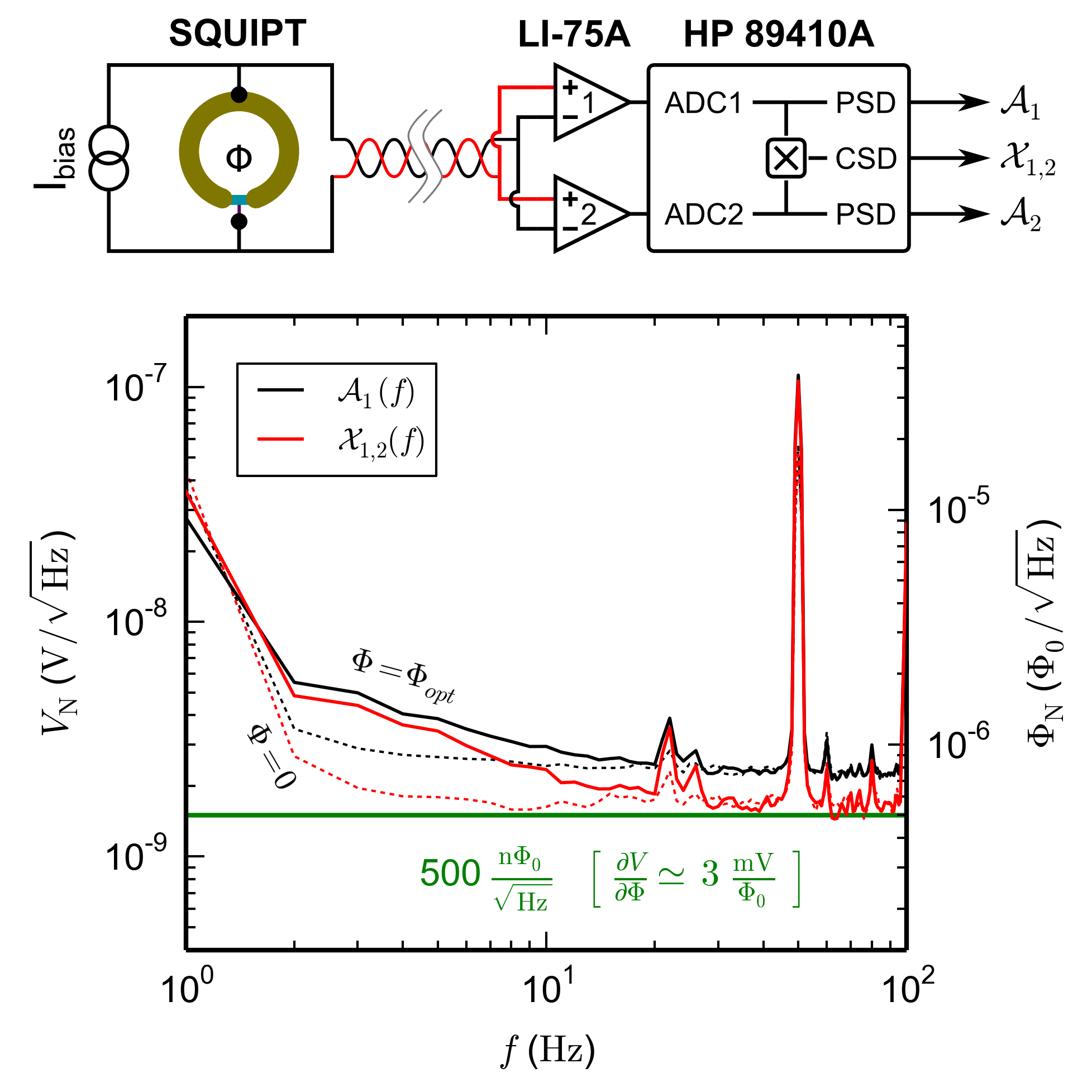}
\end{center}
\caption{
Noise level measurement in a current bias mode at $240\, \mathrm{mK}$. The schematic displayed on top summarizes the measurement
setup, in which the signal from the SQUIPT is split and fed symmetrically into two battery-powered
differential voltage preamplifiers (NF Corporation model LI-75A). The outputs from the preamplifiers
are independently sampled by two analog to digital converters (ADC) in a digital spectrum analyzer
(HP model 89410A), which computes the power spectral density (PSD, $\mathcal{A}_i$, black traces) for each ADC channel as
well as the cross spectral density (CSD, $\mathcal{X}_{1,2}$, red traces) between the two. The resulting data is expressed in 
voltage noise referred to the input of the preamplifiers ($V_N$, in $\mathrm{V/\sqrt{Hz}}$ units).
Continuous lines indicate traces recorded at optimal current and flux bias,
where the transfer function is maximal ($3\, \mathrm{mV}/\Phi_0$). The right vertical axis shows
the values for the magnetic flux resolution ($\Phi_N$) under these conditions. 
The $500\, \mathrm{n\Phi_0/\sqrt{Hz}}$ white noise level is shown as a green horizontal line. 
Dotted lines indicate control traces recorded with optimal current bias at $\Phi = 0$ 
(zero transfer function and comparable differential resistance).
}
\label{fig:fig5}
\end{figure}

Figure~\ref{fig:fig5} shows an assessment of the noise performance of the SQUIPT as a magnetic flux sensor 
obtained at $240\, \mathrm{mK}$ by measuring the power spectral density (PSD) of voltage fluctuations
recorded at the output of battery powered differential voltage preamplifiers (NF Corporation model LI-75A).
Two identical preamplification units are connected to the two independent analog to digital converter (ADC) channels
of a spectrum analyzer (HP model 89410A), which computes the PSD of each channel as well as the cross-correlated spectral
density (CSD) between the two. The latter quantifies the amount of noise which shows as correlated in the two
ADC channels, and sets an upper limit to the estimate of the intrinsic noise figures for the measurement setup.
The SQUIPT device is operated in the current-bias mode with $\mathrm{I_{bias}} = 435\, \mathrm{pA}$.
The spectral densities (both PSD and CSD) are expressed in amplitude units 
($V_N$, in $\mathrm{V/\sqrt{Hz}}$); the expected bandwidth of the measurement setup 
($\simeq 20\, \mathrm{Hz}$ when tuned for high sensitivity) is here limited by 
the significant capacitance of the filtered measurement lines ($\simeq 90\, \mathrm{nF}$).

The continuous-line traces in Fig.~\ref{fig:fig5} were acquired with the device tuned for 
maximum sensitivity ($| \partial V / \partial \Phi |_{max} = 3\, \mathrm{mV}/\Phi_0 $ and 
$\Phi_{opt} = 0.495\, \Phi_0$).
In these conditions, besides some spurious noise peaks, the input-referred white noise level
 for the preamplifiers (black trace) 
approaches the nominal limit for this model ($2\,\mathrm{nV/\sqrt{Hz}}$), 
while the cross-correlated white noise level 
(red trace) reaches values as low as $1.5\, \mathrm{nV/\sqrt{Hz}}$.  
Control traces, shown in Fig.~\ref{fig:fig5} as dotted lines,
were acquired with $\mathrm{I_{bias}} = 435\, \mathrm{pA}$ but zero magnetic flux 
(and hence zero transfer function, yet similar output differential resistance). 
They differ from the maximum-sensitivity traces for the absence 
in the $2-20\,\mathrm{Hz}$ frequency range of a 1/f slope
whose level (assuming a field-to-voltage coefficient 
$A_{eff} | \partial V / \partial \Phi |_{max} \simeq 4.8\, \mathrm{V/T}$)
is consistent with the expected magnetic low-frequency noise found in unshielded rooms in urban environment 
(typically in the $0.1 - 1\, \mathrm{nT/\sqrt{Hz}}$ range at $10\,\mathrm{Hz}$ \cite{clarke2006squid}).

The white-noise floor displayed in the CSD traces is significantly lower than the corresponding levels from
the single-channel PSD, meaning that the room-temperature preamplification stage is here limiting the noise performance
of the measurement setup. The cross-correlated voltage white-noise floor
sets an upper limit to the magnetic flux resolution achievable by our measurement setup,
$$\Phi_N = \frac{V_N}{|\partial V / \partial \Phi| } \simeq 500 \, \mathrm{n\Phi_0/\sqrt{Hz}} \, .$$
In spite of the relatively simple measurement equipment used, this noise figure is already comparable with  
state-of-the-art SNS SQUIDs interferometers equipped with custom cryogenic preamplification readout systems \cite{nagel2011superconducting,wolbing2013superconducting}.

% Conclusions / outlook
Significant improvement in terms of magnetic flux resolution would be possible with the adoption of a cryogenic preamplification to reduce the measurement noise down to $0.3\,\mathrm{nV}/\sqrt{\mathrm{Hz}}$ \cite{beev2013fully}: assuming noise figures still dominated by the readout \cite{giazotto2011hybrid}, we can expect a 5-fold increase in flux sensitivity making the present SQUIPT technology comparable to the $50\,\mathrm{n}\Phi_0/\sqrt{\mathrm{Hz}}$ resolution obtained in state-of-the-art nanoSQUIDs based on lead \cite{vasyukov2013scanning}.
 Another path is to exploit wide-gap superconductors \cite{garcia2009josephson,ronzani2013microsuperconducting,ronzani2014balanced} (such as vanadium or niobium) as proximizing elements to boost the responsivity figures.
The choice of the abovementioned wide-gap superconductors
is also a prerequisite for withstanding the sizeable magnetic field that would be necessary for 
pushing the spin resolution in nanoscale loop SQUIPTs beyond the current state-of-the-art level \cite{vasyukov2013scanning}. % Zeldov
In this respect, under optimal coupling \cite{gallop2002miniature}, 
we estimate the spin resolution of our SQUIPT $S_N = r \Phi_{N} / \pi \upmu_0 \upmu_B \approx 24\,\upmu_B/\mathrm{\sqrt{Hz}}$,
where $r$ is the SQUIPT effective radius, $\upmu_0$ is the vacuum permeability and $\upmu_B$ is the Bohr magneton.

In summary, we have presented a SQUIPT device implemented with a fabrication protocol based on the mature Al-Cu technology, 
whose design successfully resolves the limit of the incomplete phase-driven modulation of the proximity-induced minigap.
The reduced spatial extent of our SQUIPT geometry (in the micrometer range), 
combined with ultra-low power ($\approx 100\,\mathrm{fW}$)
suggest the study of the magnetic degrees of freedom of small spin populations as a possible application 
for this mesoscopic interferometer.
We conclude by emphasizing that the improvement in performance achieved in just four years 
after the initial SQUIPT proof-of-concept demonstration
put this device class already on par with the fifty-years-old SQUID technology \cite{clarke2006squid,jaklevic1964quantum}.

\begin{acknowledgments}
	The authors acknowledge M. J. Mart\'inez-P\'erez for her contribution at the early stage of this
	project and for valuable discussions.  
	The Marie Curie Initial Training Action (ITN) Q-NET 264034,
	the Italian Ministry of Education, University and Research (MIUR) 
	through the program FIRB-RBFR1379UX
        and the European Research Council under the European Union's Senventh Framework
        Programme (FP7/2007-2013)/ERC grant agreement No. 615187-COMANCHE
	are acknowledged for partial financial support. 
	C.A. thanks the Tuscany Region for funding his fellowship via the 
	CNR joint project ``PROXMAG''.
\end{acknowledgments}

\end{document}